\documentclass[conference,10pt]{IEEEtran}

\usepackage[T1]{fontenc} %
\usepackage[utf8]{inputenc} %
\usepackage[english]{babel} %
\usepackage{scalefnt} %
\usepackage{amsmath}
\usepackage{amsthm}
\usepackage{amsfonts}
\usepackage{colonequals} %
\usepackage[cmintegrals]{newtxmath} %
\usepackage{mathtools} %
\usepackage{thmtools} %
\usepackage{stmaryrd} %
\usepackage{bm} %
\usepackage{siunitx} %
\usepackage{microtype} %
\usepackage{fnpct} %
\usepackage{csquotes} %
\usepackage[inline]{enumitem} %
\usepackage{tabularx} %
\usepackage{booktabs} %
\usepackage[hidelinks]{hyperref} %
\usepackage[capitalize,nameinlink]{cleveref} %
\usepackage{xcolor} %
\usepackage{graphicx} %
\usepackage{float}  %
\usepackage{dblfloatfix} %
\usepackage{xspace} %
\usepackage{multirow} %
\usepackage{makecell} %
\usepackage{titlesec} %

\usepackage[
    backend=biber,
    style=ieee,
    citestyle=ieee-comp, %
    sortlocale=en_US,
    sortcites=true,
    url=true,
    eprint=true,
    giveninits=false,
    dashed=false,
    minnames=1,
    maxnames=5
]{biblatex}
\AtEveryBibitem{%
    \clearfield{editor}%
    \ifentrytype{inproceedings}{%
        \clearfield{publisher}%
    }{}
    \clearfield{series}%
    \clearfield{isbn}%
    \clearfield{issn}%
    \clearfield{day}%
    \clearfield{month}%
    \clearfield{place}%
    \clearlist{location}%
    \clearfield{pages}%
    \clearfield{primaryClass}%
    \clearfield{volume}%
    \clearfield{number}%
    \clearfield{note}%
    \clearfield{abstract}%
    \clearfield{keywords}%
    \clearfield{file}%
    \clearfield{language}%
    \ifentrytype{software}{%
    }{%
        \clearfield{url}%
        \clearfield{urlyear}%
        \clearfield{urldate}%
    }%
}
\newbibmacro{string+doi}[1]{%
    \iffieldundef{doi}{%
        \iffieldundef{url}{#1}{\href{\thefield{url}}{#1}}%
    }{\href{http://dx.doi.org/\thefield{doi}}{#1}}}
\DeclareFieldFormat{title}{\usebibmacro{string+doi}{\mkbibemph{#1}}}
\DeclareFieldFormat[article]{title}{\usebibmacro{string+doi}{\mkbibquote{#1}}}
\DeclareFieldFormat[inproceedings]{title}{\usebibmacro{string+doi}{\mkbibquote{#1}}}
\DeclareFieldFormat[incollection]{title}{\usebibmacro{string+doi}{\mkbibquote{#1}}}
\DeclareSourcemap{
    \maps[datatype=bibtex]{
        \map{
            \step[fieldsource=doi, match=\regexp{.+/arXiv\..+}, final]
            \step[fieldsource=eprint, match=\regexp{.+}, final]
            \step[fieldset=doi, null]
        }
        \map{
            \step[fieldsource=doi, match=\regexp{.+}, final]
            \step[fieldset=eprint, null]
            \step[fieldset=archiveprefix, null]
            \step[fieldset=eprinttype, null]
        }
    }
}
\DeclareBibliographyDriver{misc}{%
    \usebibmacro{bibindex}%
    \usebibmacro{begentry}%
    \usebibmacro{author/editor}%
    \setunit{\labelnamepunct}%
    \usebibmacro{title}%
    \newunit%
    \printlist{publisher}%
    \newunit%
    \printfield{year}%
    \usebibmacro{finentry}%
}
\makeatletter
\long\def\@makecaption#1#2{
    \ifx\@captype\@IEEEtablestring%
        \footnotesize\bgroup\par\centering\@IEEEtabletopskipstrut{\normalfont\footnotesize {#1.}\nobreakspace\scshape #2}\par\addvspace{0.5\baselineskip}\egroup%
        \@IEEEtablecaptionsepspace%
    \else
        \@IEEEfigurecaptionsepspace%
        \setbox\@tempboxa\hbox{\normalfont\footnotesize {#1.}\nobreakspace#2}%
        \ifdim\wd\@tempboxa>\hsize%
            \setbox\@tempboxa\hbox{\normalfont\footnotesize {#1.}\nobreakspace}%
            \parbox[t]{\hsize}{\normalfont\footnotesize \noindent\unhbox\@tempboxa#2}%
        \else
            \hbox to\hsize{\normalfont\footnotesize\hfil\box\@tempboxa\hfil}%
        \fi
    \fi
}
\makeatother
\makeatletter
\let\MYcaption\@makecaption%
\makeatother
\usepackage{subcaption} %
\makeatletter
\let\@makecaption\MYcaption%
\makeatother
\graphicspath{ {figures/} }

\titlespacing{\section}{0pt}{7pt}{5pt}
\titlespacing{\subsection}{0pt}{6pt}{3pt}
\setlist[enumerate]{label = (\arabic*)} %
\declaretheoremstyle[
    bodyfont=\itshape,%
    spaceabove=2pt,%
    spacebelow=2pt,%
]{example-style}
\declaretheorem[
    name=Example,%
    style=example-style,%
    numbered=unless unique,%
]{example}

\crefname{section}{Sec.}{Sec.}
\Crefname{section}{Section}{Sections}
\crefname{example}{Exp.}{Exp.}
\Crefname{example}{Example}{Examples}
\newcolumntype{R}{>{\raggedleft\arraybackslash}X}
\newcolumntype{C}{>{\centering\arraybackslash}X}

\newcommand{\ie}{i.\,e.\nolinebreak\@\xspace}

\newcommand{\eg}{e.\,g.\nolinebreak\@\xspace}
\newcommand{\Eg}{E.\,g.\nolinebreak\@\xspace}

\newcommand{\sci}[2]{$#1\!\times\!10^{#2}$} %
\renewcommand{\phi}{\varphi}

\definecolor{TUM_blue}{RGB}{0,101,189}
\colorlet{TUM_black}{black}
\colorlet{TUM_white}{white}
\definecolor{TUM_darkblue}{RGB}{0,82,147}
\colorlet{TUM_darkblue100}{TUM_darkblue}
\colorlet{TUM_darkblue80}{TUM_darkblue100!80}
\colorlet{TUM_darkblue50}{TUM_darkblue100!50}
\colorlet{TUM_darkblue20}{TUM_darkblue100!20}
\definecolor{TUM_verydarkblue}{RGB}{0,51,89}
\colorlet{TUM_verydarkblue100}{TUM_verydarkblue}
\colorlet{TUM_verydarkblue80}{TUM_verydarkblue100!80}
\colorlet{TUM_verydarkblue50}{TUM_verydarkblue100!50}
\colorlet{TUM_verydarkblue20}{TUM_verydarkblue100!20}
\colorlet{TUM_darkgrey}{TUM_black!80}
\colorlet{TUM_grey}{TUM_black!50}
\colorlet{TUM_lightgrey}{TUM_black!20}
\definecolor{TUM_beige}{RGB}{218,215,203}
\definecolor{TUM_orange}{RGB}{227,114,34}
\definecolor{TUM_green}{RGB}{162,173,0}
\definecolor{TUM_verylightblue}{RGB}{152,198,234}
\definecolor{TUM_lightblue}{RGB}{100,160,200}

\title{Scalable Circuit Cutting: A Framework for Combined Gate and Wire Cuts Using Gate Groups}

\author{
    \IEEEauthorblockN{
        Fiona Jiali Fröhler\IEEEauthorrefmark{1},
        Yannick Stade\IEEEauthorrefmark{1},
        Christian Ufrecht, %
        Daniel D. Scherer\IEEEauthorrefmark{3},
        and Robert Wille\IEEEauthorrefmark{1}\IEEEauthorrefmark{4}
    }
    \IEEEauthorblockA{\IEEEauthorrefmark{1}%
    Chair for Design Automation,
        Technical University of Munich,
        Munich, Germany
    }
    \IEEEauthorblockA{\IEEEauthorrefmark{3}%
    Fraunhofer IIS, Fraunhofer Institute for Integrated Circuits IIS,
        Nuremberg, Germany
    }
    \IEEEauthorblockA{\IEEEauthorrefmark{4}%
    MQSC,
        Garching near Munich, Germany
    }
    \{%
    fiona.froehler, %
    yannick.stade, %
    robert.wille\}@tum.de, 
    christian.ufrecht@gmx.de, 
    daniel.scherer2@iis.fraunhofer.de\\
    \href{https://www.cda.cit.tum.de/research/quantum}{www.cda.cit.tum.de/research/quantum}
}
\hypersetup{ %
    pdftitle={Heuristic Graph Partitioning for Circuit Cutting},
    pdfsubject={IEEE International Conference on Quantum Computing and Engineering 2026},
    pdfauthor={
        Fiona Fröhler,
        Yannick Stade,
        Robert Wille
    }
}

\begin{document}

    \maketitle

    \begin{abstract}
        Quantum circuit cutting enables the execution of large circuits on devices with a limited number of qubits by partitioning circuits into independent subcircuits.
        However, this introduces a sampling overhead, which grows exponentially with the number of cuts, rendering the choice of cut placements critical for practical circuit cutting.
        Determining optimal cut placements remains computationally challenging, particularly as circuits grow in size.
        Additionally, most existing circuit cutting frameworks treat gate and wire cuts independently.
        Moreover, approaches combining both cutting, however, do not take advantage of joint cutting, \ie, identifying common gate groups and cutting them jointly to reduce overhead.
        This work presents a unified framework that, for the first time, combines gate and wire cutting with joint cutting in a scalable manner.
        To this end, we propose a novel \mbox{gate-group-aware} technique to further reduce sampling overhead.
        By formulating the cut placement problem as a scalable graph partitioning task, the proposed method efficiently identifies \mbox{high-quality} cut placements for large circuits, also providing diagnostic feedback on whether circuits are suitable for cutting.
    \end{abstract}


    \section{Introduction}\label{sec:introduction}

Quantum algorithms such as Shor's factoring algorithm~\parencite{shor_algorithms_1994} and Grover's search~\parencite{grover_fast_1996} offer the promise of significant computational advantages over classical methods.
However, realizing these speedups on current quantum hardware remains a significant challenge.
Today's quantum processors are constrained by a limited number of qubits, high error rates, and restricted connectivity—characteristics that collectively define the \emph{Noisy \mbox{Intermediate-Scale} Quantum} (NISQ) era~\parencite{preskill_nisq_2018}.
These limitations make it impractical to execute large, complex quantum circuits on a single device.

To overcome this issue, \emph{circuit cutting}\textemdash{}also called \emph{circuit knitting}—has emerged as a practical strategy for decomposing large quantum circuits into smaller subcircuits that can be executed independently with only classical communication channels between devices~\parencite{hofmann_how_2009,peng_simulating_2020,mitarai_constructing_2021}.
This approach has been experimentally realized on real hardware~\parencite{ying_experimental_2023, singh_experimental_2024, carrera_vazquez_combining_2024,herzog_improving_2024} and does not require quantum communication channels, unlike \emph{Distributed Quantum Computing}~(DQC), making it attractive for \mbox{near-term} systems.

However, circuit cutting comes with a fundamental \mbox{trade-off}: the sampling overhead grows exponentially with the number of cuts, typically limiting practical application to around five to eight cuts~\parencite{brandhofer_optimal_2024}.
This makes \emph{cut placement}\textemdash{}the process of identifying where to cut a circuit\textemdash{}critical, as \mbox{high-quality} placements directly reduces the required number of cuts.

The two main strategies for cut placement are \emph{gate cuts} and \emph{wire cuts}.
Most prior frameworks focus on one strategy in isolation, potentially missing opportunities for more flexible and efficient cut locations.
Furthermore, recent work has shown that jointly cutting multiple gates together as a group can substantially reduce the sampling overhead compared to individual cuts~\parencite{piveteau_circuit_2024}\textemdash{}yet this technique is rarely combined with traditional cutting strategies.

This work presents a unified framework that combines gate and wire cutting with joint cutting within a single method.
By leveraging all three techniques, the approach achieves better scalability and enables circuit cutting across a broader class of circuits.
The framework uses graph partitioning heuristics to efficiently identify promising cut locations, naturally balancing solution quality and computational speed, thereby making circuit cutting more practical and suitable for integration into larger quantum software stacks such as the MQSS~\parencite{mqss}. %

    \begin{figure*}[htbp]
    \centering
    \begin{subfigure}[t]{0.5\linewidth}
        \centering
        \includegraphics[trim=0 0 23pt 23pt,clip,width=0.79\linewidth]{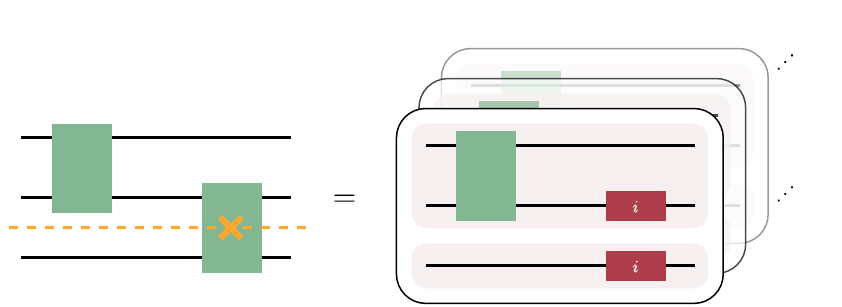}
        \vspace{1pt}
        \caption{Example of gate cutting, where \mbox{multi-qubit} gates are replaced with a classical \mbox{quasi-probability} decomposition over \mbox{single-qubit} gates labelled by index $i$.}\label{fig:gate-cut}
    \end{subfigure}
    \hspace{0.005\linewidth}
    \begin{subfigure}[t]{0.48\linewidth}
        \centering
        \includegraphics[trim=0 0 23pt 23pt,clip,width=0.81\linewidth]{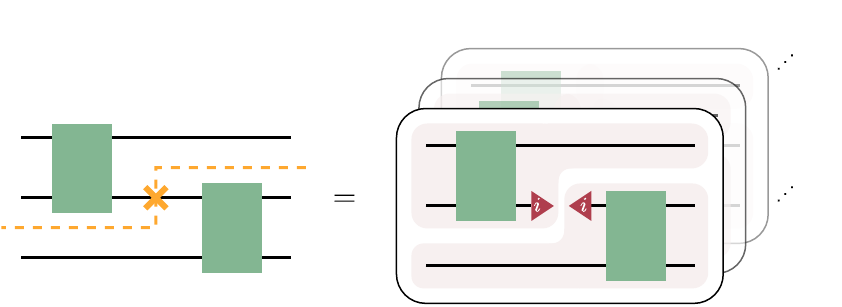}
        \vspace{1pt}
        \caption{Example of a wire cut, where the flow of quantum information is interrupted along a qubit wire by measuring and preparing the qubit state.}\label{fig:wire-cut}
    \end{subfigure}

    \vspace{5pt}
    
    \begin{subfigure}[t]{\linewidth}
        \centering
        \includegraphics[trim=7pt 0 0 10pt,clip,width=\linewidth]{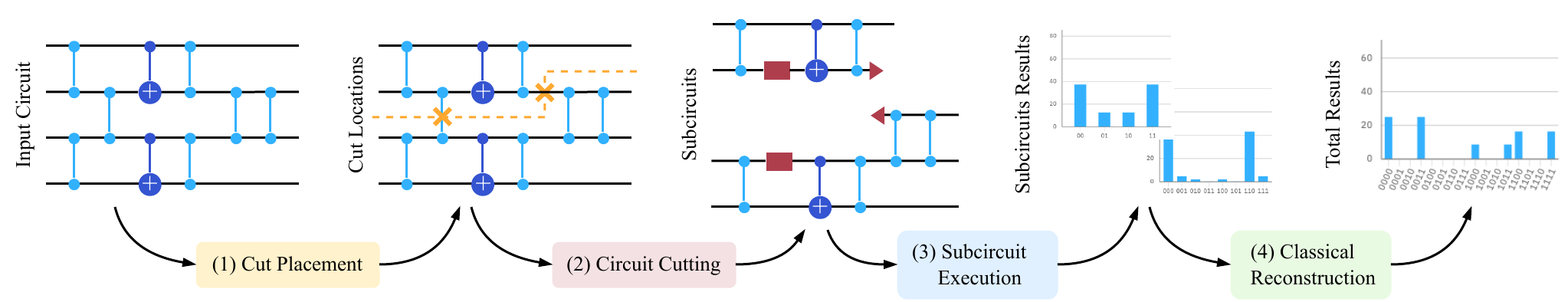}
        \caption{Schematic workflow of a general circuit cutting procedure. This work focuses on the first step\textemdash{}cut placement\textemdash{}highlighted in yellow: identifying a \mbox{high-quality} set of cut locations to partition the circuit.}\label{fig:workflow}
    \end{subfigure}
    \caption{Different kinds of quantum circuit cuts and the circuit cutting workflow.}\label{fig:circuit-cutting}
    \vspace*{-12pt}
\end{figure*}

\section{Preliminaries}\label{sec:background}

To keep this paper \mbox{self-contained}, this section provides the necessary background for the circuit cutting problem addressed in this work.
It reviews the conceptual foundations of quantum circuit cutting and existing approaches to cut placement.

\subsection{Quantum Circuit Cutting}

Quantum circuit cutting~\parencite{hofmann_how_2009,peng_simulating_2020,mitarai_constructing_2021} enables the execution of large quantum circuits that do not fit as a whole on available hardware with a limited number of qubits.
To this end, it partitions circuits into smaller subcircuits, each executable on the available hardware.

Circuit cutting can be realized through two approaches.
\emph{Gate cuts}~\parencite{mitarai_constructing_2021,hofmann_how_2009,mitarai_overhead_2021,ufrecht_cutting_2023,ufrecht_optimal_2024,piveteau_circuit_2024,schmitt_cutting_2025,harrow_optimal_2025} remove \mbox{multi-qubit} gates and replace them with local gates as shown in~\cref{fig:gate-cut}.
\emph{Wire cuts}~\parencite{peng_simulating_2020,brenner_optimal_2023,harada_doubly_2024,pednault_alternative_2023,lowe_fast_2023,harrow_optimal_2025} break the flow of quantum information along a qubit wire as illustrated in~\cref{fig:wire-cut}.
Both approaches cut the circuit into smaller subcircuits, each satisfying the hardware's qubit count.
Formally, this is described by a \mbox{quasi-probability} decomposition of quantum channels~\parencite{pashayan_estimating_2015,piveteau_quasiprobability_2022}.

However, this gained flexibility comes at a price: every subcircuit must be executed multiple times to be able to reconstruct the results of the entire circuit from the executions of the subcircuits.
This \mbox{so-called} \emph{sampling overhead} scales exponentially with the number of cuts.
Specifically, the overhead for estimation of the expectation value of an observable scales as $\kappa^{2n}$ for $n$ independent cuts, where $\kappa$ is defined through the \mbox{quasi-probability} decomposition.
For gate cuts, $\kappa_g = 3$ for CNOT gates~\parencite{hofmann_how_2009}, $\kappa_g = 1 + 2|\sin(\theta)|$ for \mbox{two-qubit} rotation gates~\parencite{mitarai_constructing_2021,piveteau_circuit_2024} and $\kappa_g = 7$ for swap gates~\parencite{mitarai_overhead_2021,piveteau_circuit_2024}.
For wire cuts, the optimal overhead is $\kappa_w = 4$ (without classical communication)~\parencite{peng_simulating_2020}, \ie, for one wire cut, each subcircuit needs $4^2 = 16$ times more samples.

The resulting circuit cutting workflow is depicted in~\cref{fig:workflow} and consists of four main phases:
(1)~\emph{cut placement} to determine an optimal cutting strategy that minimizes sampling overhead while ensuring subcircuits fit on available hardware,
(2)~\emph{circuit cutting} where the individual subcircuits are constructed based on the cut locations and the chosen cutting strategy,
(3)~\emph{subcircuit execution} where each subcircuit is executed multiple times independently, %
and (4)~\emph{classical reconstruction} to recombine outputs via \mbox{post-processing}.

\subsection{Combined Cutting}

The exponential scaling of the sampling overhead limits practicality: typically only five to eight cuts remain feasible before overhead becomes prohibitive~\parencite{brandhofer_optimal_2024}.
Consequently, minimizing the number of cuts is essential for practical circuit cutting.
To achieve this, considering gate and wire cuts simultaneously can yield subcircuits with significantly fewer total cuts, directly reducing sampling overhead.

\begin{figure*}[htbp]
    \centering
    \begin{subfigure}{0.31\linewidth}
        \centering
        \includegraphics[trim = 2mm 0 0mm 0,width=0.9\linewidth]{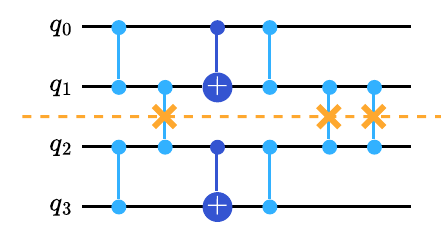}
        \caption{Optimal gate cut, $\kappa = 3 \cdot 3 \cdot 3 = 27$}\label{fig:example-gate}
    \end{subfigure}
    \hspace{0.01\linewidth}
    \begin{subfigure}{0.31\linewidth}
        \centering
        \includegraphics[trim = 2mm 0 0mm 0,width=0.9\linewidth]{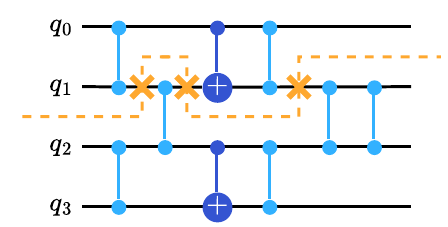}
        \caption{Optimal wire cut, $\kappa = 4 \cdot 4 \cdot 4 = 64$}\label{fig:example-wire}
    \end{subfigure}
    \hspace{0.01\linewidth}
    \begin{subfigure}{0.31\linewidth}
        \centering
        \includegraphics[trim = 2mm 0 0mm 0,width=0.9\linewidth]{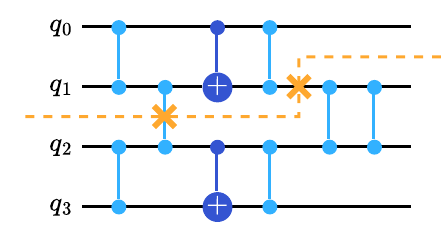}
        \caption{Optimal combined cut, $\kappa = 3 \cdot 4 = 12$}\label{fig:example-combined}
    \end{subfigure}
    \vspace{-5pt}
    \caption{Example illustrating the advantage of combined cutting. While pure gate or wire cuts may be optimal within their own categories, a unified strategy can identify mixed solutions that significantly reduce total sampling overhead.}\label{fig:example}
    \vspace*{-12pt}
\end{figure*}

\begin{example}\label{ex:combined-cut}
    \cref{fig:example} depicts three different cut placements for the same circuit, in which single-qubit gates were omitted.
    The optimal gate cut of the given circuit depicted in~\cref{fig:example-gate} yields $\kappa = 27$, and the optimal wire cut shown in~\cref{fig:example-wire} yields $\kappa = 64$.
    However, a unified strategy, as illustrated in~\cref{fig:example-combined}, achieves $\kappa = 12$ by leveraging the strengths of both approaches.
\end{example}

As~\cref{ex:combined-cut} demonstrates, the individual strategies alone can be suboptimal depending on the circuit topology. %
Consequently, a comprehensive strategy should consider both gate and wire cuts simultaneously.

\subsection{Joint Cutting}

\begin{figure}[htbp]
    \vspace{-4pt}
    \centering
    \begin{subfigure}[t]{0.48\linewidth}
        \centering
        \includegraphics[trim = 0mm 0 -7mm 0,width=0.74\linewidth]{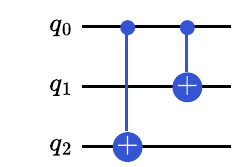}
        \caption{Adjacent controls: \mbox{Two-qubit} gates adjacent on a shared control qubit. The sampling overhead for cutting at any position scales with $\kappa=3$.}\label{fig:gate-group-cascade-control}
    \end{subfigure}
    \hspace{0.01\linewidth}
    \begin{subfigure}[t]{0.48\linewidth}
        \centering
        \includegraphics[trim = 0mm 0 -7mm 0,width=0.74\linewidth]{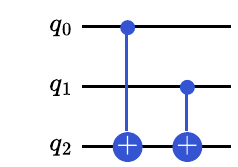}
        \caption{Adjacent targets: \mbox{Two-qubit} gates with the same effect adjacent on a shared target qubit. The sampling overhead for cutting at any position scales with $\kappa=3$.}\label{fig:gate-group-cascade-target}
    \end{subfigure}
    
    \vspace{-5pt}

    \begin{subfigure}[t]{0.48\linewidth}
        \centering
        \includegraphics[trim = 0mm 0 -7mm 0,width=0.74\linewidth]{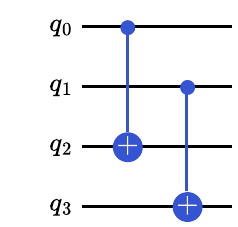}
        \caption{Parallel gates: \mbox{Two-qubit} gates acting in parallel on disjoint qubits. Joint cutting can reduce the overhead for cutting $n_g$ gates to $\kappa = 2^{n_g+1} - 1$, \eg, for two gates $\kappa=7$ instead of $\kappa=9$.}\label{fig:gate-group-cascade-parallel}
    \end{subfigure}
    \hspace{0.01\linewidth}
    \begin{subfigure}[t]{0.48\linewidth}
        \centering
        \includegraphics[trim = 0mm 0 -7mm 0,width=0.74\linewidth]{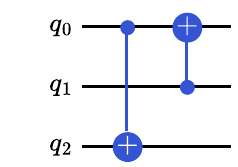}
        \caption{Other groupings: More complex constellations, \eg\ gates adjacent on the same qubit but not a cascade. \Eg\ jointly cutting both gates here yields $\kappa=7$, whereas individually cutting them yields $\kappa=9$.}\label{fig:gate-group-cascade-example}
    \end{subfigure}
    \caption{Examples of different types of gate groups and their associated sampling overheads for various cut placements.}\label{fig:gate-group-cascade}
\end{figure}

Beyond considering both cut types, further research explored \emph{joint cutting}, which involves cutting multiple gates together or multiple wires together as a group, rather than independently.
This can further reduce sampling overhead~\parencite{ufrecht_optimal_2024,piveteau_circuit_2024,schmitt_cutting_2025,harrow_optimal_2025,brenner_optimal_2023,harada_doubly_2024,lowe_fast_2023}.
For gates, this advantage can only be exploited for certain compositions, which we refer to as \emph{gate groups}.
These gate groups are illustrated in~\cref{fig:gate-group-cascade} and classified by their structure:

\begin{itemize}
    \item \textbf{Adjacent controls} (see~\cref{fig:gate-group-cascade-control}): Two or more \mbox{two-qubit} gates that share a control qubit and are adjacent along the control line, form a \mbox{so-called} \emph{cascade} on the controls. For CNOT gates, the sampling overhead for cutting the gate group at any position scales with $\kappa=3$.
    \item \textbf{Adjacent targets} (see~\cref{fig:gate-group-cascade-target}): Similarly, \mbox{two-qubit} gates that share a target qubit and are adjacent along the target, form a \emph{cascade} on the targets, with $\kappa=3$ for any cut position. However, the gates must be of the same type.
    \item \textbf{Parallel gates} (see~\cref{fig:gate-group-cascade-parallel}): $n_g$ \mbox{two-qubit} gates act in parallel on disjoint sets of qubits. Here, joint cutting can reduce the overhead below the product of individual cuts, with the overhead for jointly cutting $n_g$ CNOT gates scaling with $\kappa = 2^{n_g+1} - 1$ instead of $\kappa = 3^{n_g}$~\parencite{ufrecht_optimal_2024,schmitt_cutting_2025,harrow_optimal_2025}.
    \item \textbf{Other groupings} (see~\cref{fig:gate-group-cascade-example}): More complex constellations, such as gates adjacent on the same qubit but not being a cascade, or different gates being adjacent on their target, can also form gate groups with potentially reduced overhead.
\end{itemize}

Similarly, wires can be jointly cut to reduce sampling overhead below the product of individual wire cuts, however this requires classical communication between subcircuits~\mbox{\parencite{brenner_optimal_2023,harada_doubly_2024,lowe_fast_2023}}.

\subsection{Existing Approaches}

Most existing approaches to cut placement focus on either gate cuts or wire cuts separately.
Wire cutting has established optimal methods, such as CutQC~\parencite{tang_cutqc_2021}, as well as more recent heuristics~\parencite{tang_scaleqc_2022,kan_fitcut_2024,basu_fragqc_2024}.
Gate cut placement is less studied, but can leverage techniques from Distributed Quantum Computing~(DQC)~\parencite{zomorodi-moghadam_optimizing_2018, andres-martinez_automated_2019, daei_optimized_2020, davarzani_dynamic_2020, chen_distributed_2025, cambiucci_spatial_2025}.

However, as demonstrated in~\cref{ex:combined-cut}, separate consideration of gate and wire cuts prevents finding optimal decompositions for circuits that would benefit from combined strategies.
So far, only one method solves the cut placement problem optimally for combined gate and wire cuts~\parencite{brandhofer_optimal_2024}. The corresponding \mbox{SMT-based} approach, while exact, is limited to small circuits (up to 1 hour for \mbox{40-qubit} circuits with 51 gates).

Recent heuristics address this gap with varied strategies.
Pawar et al.~\parencite{pawar_qrcc_2025} combine an \mbox{ILP-based} cut finder with qubit reuse, demonstrating improvements over CutQC~\parencite{tang_cutqc_2021} while facing scalability challenges.
The Qiskit circuit cutting module~\parencite{qiskit-addon-cutting} and Nakamura et al.~\parencite{nakamura_improved_2025} target \mbox{k-way} partitioning instead of bi-partitioning; the latter, though claiming superior performance over Qiskit, may produce vertical cuts that satisfy hardware constraints but incur unnecessary sampling overhead.
For DQC, Burt et al.~\parencite{burt_generalised_2024} demonstrate strong performance on structured circuits using genetic algorithms with graph partitioning.
Other DQC approaches~\mbox{\parencite{davis_towards_2023, nikahd_automated_2021, crampton_genetic_2025}} are slower or too \mbox{domain-specific} for direct application to circuit cutting.

Furthermore, a key limitation across existing approaches concerns gate groups: they only handle cascades, \ie, the cases shown in \cref{fig:gate-group-cascade-control} and \cref{fig:gate-group-cascade-target}, missing the efficiency gains achievable through general joint cutting~\parencite{andres-martinez_automated_2019, burt_generalised_2024}.

    \section{Considered Problem}\label{sec:problem}

In this work, we focus on \emph{cut placement}\textemdash{}the first step of a circuit cutting workflow as shown in~\cref{fig:workflow}.
As outlined in the previous section, existing methods typically optimize either gate or wire cuts individually and do not consider gate groups for joint cutting.
Although combining gate and wire cutting with joint cutting can significantly reduce sampling overhead compared to individual cuts.

\begin{figure}[htbp]
    \vspace*{-5pt}
    \centering
    \begin{subfigure}{0.484\linewidth}
        \centering
        \includegraphics[trim = 7mm -2mm 4mm 0,clip,width=\linewidth]{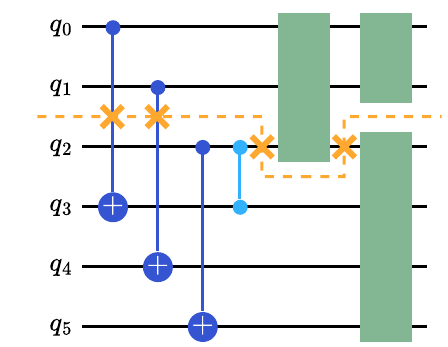}
        \vspace{-7pt}
        \caption{Optimal combined cut using individual cuts, $\kappa = 3\cdot 3\cdot 4\cdot 4=144$}\label{fig:problem-example-combined}
    \end{subfigure}
    \hspace{0.005\linewidth}
    \begin{subfigure}{0.484\linewidth}
        \centering
        \includegraphics[trim = 7mm 0 4mm 0,clip,width=\linewidth]{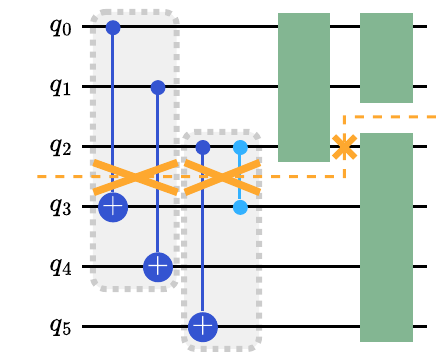}
        \vspace{-7pt}
        \caption{Optimal combined cut considering gate groups, $\kappa = (2^{2+1}-1)\cdot 3\cdot 4=84$}\label{fig:problem-example-gate-groups}
    \end{subfigure}
    \vspace{-17pt}
    \caption{Example of combined gate, wire, and joint cuts.}\label{fig:problem-example}
    \vspace*{4pt}
\end{figure}

\begin{example}
    Consider the \mbox{6-qubit} circuit shown in~\cref{fig:problem-example}.
    In~\cref{fig:problem-example-combined}, applying individual gate and wire cuts yields a sampling overhead of $\kappa = 3\cdot 3\cdot 4\cdot 4=144$.
    However, the circuit contains gate groups: two parallel gates on qubits $q_0,q_3$ and $q_1,q_4$, and a cascade on control qubit $q_2$.
    By cutting these gates jointly rather than individually, the overhead in~\cref{fig:problem-example-gate-groups} is reduced to $\kappa = (2^{2+1}-1)\cdot 3\cdot 4=84$.
    This corresponds to a 41\% reduction compared to individual cuts.
\end{example}

The example demonstrates that the existing methods leave huge potential for combining gate and wire cutting with joint cutting untapped.
Consequently, we target the following problem statement:
\begin{itemize}
    \item \textbf{Input:} A quantum circuit $C$ consisting of a set of qubits $Q$ and gates, along with possible gate groups.
    \item \textbf{Constraints:} A \mbox{hardware-imposed} limit $q_{\max}$, representing the maximum number of qubits available per subcircuit. The solution must ensure $|Q_i| \leq q_{\max}$ for each subcircuit $C_i$.
    \item \textbf{Output:} A set of \emph{cut locations} that partition the circuit into independent subcircuits respecting the qubit constraint.
    Cuts can be individual gate cuts, wire cuts, or joint cuts of groups of gates.
    \item \textbf{Objective:} Minimizing \emph{sampling overhead} $S = \kappa_g^{2n_g} \cdot \kappa_w^{2n_w} \cdot S_{group}$, where $n_g$ and $n_w$ are the numbers of gate and wire cuts, $\kappa_g$ and $\kappa_w$ result from the chosen cutting strategy, and $S_{group}$ is the overhead from joint cutting.
\end{itemize}
This work uses minimal \mbox{$\kappa$-values} that do not require classical communication between subcircuits.

Determining the optimal solution to the problem above can be formulated as a balanced graph partitioning problem, which is known to be \mbox{NP-hard}~\parencite{andreev_balanced_2004}.
This \mbox{NP-hardness} creates a fundamental barrier: exact methods cannot scale to practical qubit counts.
Aside from scalability, another critical gap remains\textemdash{}none of the existing methods for circuit cutting address this problem within a unified framework.

    \section{Proposed Solution}\label{sec:approach}

This work addresses this deficiency by proposing a novel scalable approach that combines gate and wire cuts and considers gate groups for joint cutting, allowing circuits with hundreds of qubits to benefit from synergies between these techniques.
To this end, we propose a scalable \mbox{heuristic-based} algorithm grounded in graph partitioning techniques that efficiently approximates the optimal solution while scaling to practical circuit sizes.

It follows a \emph{multilevel partitioning strategy} that progressively refines the cut placement.
The first level identifies a good gate cut (ideally close to optimal) by only focusing on the circuit's spatial structure.
The second level then refines this solution by replacing gate cuts with wire cuts, where this reduces the overall sampling overhead.
Furthermore, both steps are extended to further reduce overhead by identifying gate groups and cutting them jointly rather than individually.
Each stage is designed to maintain or improve upon prior solutions, with detailed algorithms presented later in~\cref{sec:details}.

\begin{figure}[ht]
    \vspace*{-6pt}
    \centering
    \includegraphics[width=0.8\linewidth]{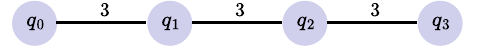}
    \vspace*{-3pt}
    \caption{Mapping circuit in~\cref{fig:example} to an interaction graph to find gate cuts.}\label{fig:interaction-graph}
\end{figure}

The first stage considers only gate cuts, simplifying the initial problem.
We construct an \emph{interaction graph} (see \cref{fig:interaction-graph}) where each qubit is a node and edges represent \mbox{two-qubit} operations between qubits.
The weight of each edge reflects how frequently two qubits interact.
By finding a balanced partition of this graph, we identify which interactions are expensive to cut and should be kept within the same partition, and which can be sacrificed.

\begin{itemize}[label={},leftmargin=0pt]
    \item \textbf{Input}: A quantum circuit with single- or \mbox{two-qubit} gates
    \item \textbf{Output}: A partition of qubits into two groups, and the number of gates cut by this partition
\end{itemize}

\begin{example}\label{ex:interaction-graph}
    Consider the interaction graph in~\cref{fig:interaction-graph} of the \mbox{4-qubit} circuit shown in~\cref{fig:example}.
    The interaction graph formalizes which qubits communicate and how intensely, revealing a chain of equally strong connections ($q_0$--$q_1$--$q_2$--$q_3$).
    By inspecting the graph, we identify the edge between $q_1$ and $q_2$ as an ideal cutting location for a balanced cut.
    Partitioning as $\{q_0, q_1\}$ vs.~$\{q_2, q_3\}$ cuts only one edge (weight~$3$), minimizing the gates affected and thus the sampling overhead.
\end{example}

\begin{figure}[t]
    \vspace*{-5pt}
    \centering
    \includegraphics[width=0.7\linewidth]{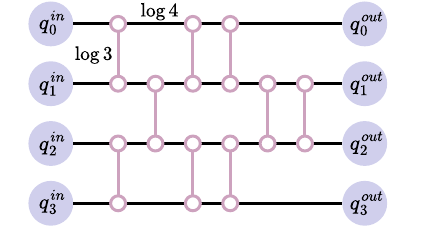}
    \vspace*{-3pt}
    \caption{Mapping circuit in~\cref{fig:example} to a circuit topology graph to find gate and wire cuts.}\label{fig:flow-graph}
\end{figure}

The second stage refines the partition by considering both gate cuts and wire cuts simultaneously.
After identifying good gate cut locations, we construct a \emph{circuit topology graph} (see \cref{fig:flow-graph}) that exposes the temporal structure of the circuit.
As depicted in~\cref{fig:flow-graph}, the circuit topology graph not only shows the number of interactions between the qubits but also the time and order of the interactions.
This allows us to selectively cut some qubit wires (in addition to or instead of certain gates) if doing so reduces overall overhead.
The key challenge is ensuring that cuts remain valid for the given qubit bounds.

\begin{itemize}[label={},leftmargin=0pt]
    \item \textbf{Input}: Quantum circuit and initial partition from the first stage
    \item \textbf{Output}: Refined partition using both gate and wire cuts
\end{itemize}

\begin{example}\label{ex:flow-graph}
    Consider the circuit topology graph in~\cref{fig:flow-graph} of the \mbox{4-qubit} circuit shown in~\cref{fig:example}.
    Then, starting with the gate cut identified in the first stage, we can explore beneficial wire cuts, \eg, cutting a wire instead of two gates.
    This stage should ideally result in the combined cut shown in~\cref{fig:example-combined} that achieves the lowest sampling overhead.
\end{example}

This \mbox{two-level} approach guarantees that we never perform worse than pure gate cutting alone, while capturing synergies between gate and wire cuts.

While the presented approach already constitutes a valid solution to the problem, some resulting cuts can be suboptimal.
To this end, we utilize gate groups.
By cutting gate groups\textemdash{}rather than individual gates\textemdash{}the sampling overhead can be further reduced.
Gate groups are incorporated as \mbox{partition-dependent} cost functions that compute sampling overhead based on the qubit partition, rather than using fixed costs, which would be the case if they were modeled with hyperedges.

Gate groups are incorporated by augmenting the graph with \mbox{partition-dependent} cost functions that accurately reflect the overhead reduction when cutting gates jointly rather than individually. This allows both stages to identify cuts that exploit these grouping opportunities.

\begin{example}\label{ex:gate-grouping}
    Revisiting the \mbox{6-qubit} circuit in~\cref{fig:problem-example}, the circuit topology graph alone identifies cuts that yield sampling overhead \mbox{$\kappa = 144$} (as shown in the problem section).
    By recognizing the parallel gates and cascade structure as exploitable groups, the algorithm can identify the same cuts but account for the efficiency gained by joint cutting, yielding the improved solution with overhead \mbox{$\kappa = 84$} referenced in~\cref{fig:problem-example-gate-groups}.
\end{example}

\begin{figure}[ht]
    \vspace*{-4pt}
    \centering
    \includegraphics[width=0.7\linewidth]{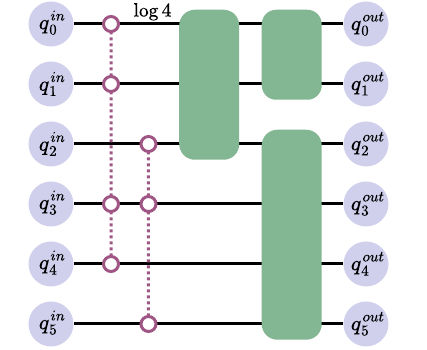}
    \vspace*{-3pt}
    \caption{Mapping circuit in~\cref{fig:problem-example} to a circuit topology graph to find gate and wire cuts while considering gate groups.}\label{fig:gate-grouping}
\end{figure}

    \section{Implementation Details}\label{sec:details}

This section describes the implementation of the \mbox{multi-level} circuit cutting approach resulting from the ideas sketched above.

\subsection{Gate Cutting via Interaction Graph}\label{sec:impl-gate-cutting}

The first stage performs spatial partitioning via gate cutting by mapping the quantum circuit to an interaction graph and then applying balanced graph partitioning heuristics.

We construct an interaction graph \mbox{$G = (V, E)$} as follows:
\begin{itemize}
    \item \textbf{Vertices} ($V$): For each qubit $q_i$, create a node
    \item \textbf{Edges} ($E$): For each \mbox{two-qubit} gate operating on qubits $q_i$ and $q_j$, add or increment the edge $(q_i, q_j)$ in the graph
    \item \textbf{Edge weights}: The weight $w(i,j)$ equals the total number of \mbox{two-qubit} gates between qubits $i$ and $j$
    \item \textbf{Exclusions}: \mbox{Single-qubit} gates are ignored; \mbox{multi-qubit} gates are decomposed into \mbox{two-qubit} gates
\end{itemize}

We partition the interaction graph using the \mbox{Kernighan-Lin} (KL) heuristic~\parencite{kernighan_efficient_1970}, implemented in NetworkX~\parencite{networkx}.
KL iteratively swaps pairs of vertices between partitions to minimize the total edge cut cost while maintaining balance of vertices.
Since KL is sensitive to initialization, we run the algorithm multiple times with different random initial partitions and select the partition with the minimum edge cut cost.

\subsection{Combined Cutting via Circuit Topology Graph}

To incorporate temporal information about gate order, we augment the interaction graph with a circuit topology graph $G = (V_Q \cup V_G, E_W \cup E_G)$ that represents the time dimension:
\begin{itemize}
    \item \textbf{Qubit vertices} ($V_Q$): For each qubit, create two vertices: $q_\text{in}$ (input) and $q_\text{out}$ (output)
    \item \textbf{Gate vertices} ($V_G$): For each \mbox{two-qubit} gate, create two gate vertices
    \item \textbf{Gate edges} ($E_G$): Connect the two vertices of each gate
    \item \textbf{Wire edges} ($E_W$): For each qubit, connect temporally consecutive vertices with an edge, representing the qubit's evolution through the circuit
\end{itemize}
To apply KL to the circuit topology graph, we assign weights and vertex sizes aligning with the underlying circuit constraints.
Edge weights use logarithmic scaling to ensure linear optimization corresponds to exponential overhead reduction, \ie,
\begin{align}
    w_g &= \log \kappa_g, \quad w_w = \log \kappa_w
\end{align}
where $\kappa_g \in \{3, 1 + 2|\sin(\theta)|, 7\}$ (see~\cref{sec:background}) and $\kappa_w = 4$ are the minimal sampling overheads for cutting a single \mbox{two-qubit} gate and wire, respectively.

We assign vertex sizes to enforce the balance constraint only on qubits: gate vertices have size 0, while qubit vertices have size 1.
This reflects the physical constraint that device qubit counts are fixed, whereas the number of gates is irrelevant to the partition.
We adapted the NetworkX implementation of KL to respect these vertex sizes during partitioning.

Direct partitioning of the circuit topology graph would be problematic: the graph structure is more permissive than the circuit structure.
It allows \mbox{graph-valid} partitions that violate qubit constraints (\eg, vertical cuts separating all wires at a single time step).
The \mbox{two-stage} strategy addresses this by initializing KL with the spatial partition from the first stage, which prevents such invalid configurations, then refining it by exploring wire cuts while staying within qubit bounds.
This guarantees that the final partition respects qubit constraints, never performs worse than the first stage alone, and captures synergies between gate and wire cuts.

\subsection{Extension: Gate Groups}

In the partitioning framework, we represent gate groups (detailed in~\cref{sec:background}) as \mbox{partition-dependent} cost functions.
For a given partition of qubits, the function returns the actual sampling overhead incurred by cutting that group, which may vary depending on which qubits are separated.

To incorporate gate groups into the partitioning process, we modify the KL algorithm in two ways.
First, when constructing the circuit topology or interaction graph, we remove all edges corresponding to gates within a gate group.
This prevents \mbox{double-counting} their cost, since their contribution is instead handled through the gate group cost function during the KL gain calculations.

Second, we adjust the \mbox{so-called} \mbox{D-values} and swap gains within the KL gain calculations to account for gate groups.
The \mbox{D-value} for a vertex $v$ normally represents the change in cost if $v$ moves to the opposite partition.
If $v$ participates in a gate group, we extend the \mbox{D-value} calculation to:
\begin{equation}
    D[v] = \text{ExtCost}(v) - \text{IntCost}(v) + (c_\text{current} - c_\text{moved}),
\end{equation}
where $c_\text{current}$ is the current cut cost of the gate group and $c_\text{moved}$ is the cost if $v$ were moved.
This ensures that changes in the gate group's cost are reflected in the gain calculation.

Similarly, when evaluating a swap of two vertices $a$ and $b$, we adjust the swap gain to:
\begin{equation}
    \delta = D[a] + D[b] - 2w(a,b) + (c_\text{swapped} - c_\text{current}),
\end{equation}
where $c_\text{swapped}$ is the gate group's cost if $a$ and $b$ are swapped. This accounts for any changes in the joint cutting cost resulting from the combined move.

\mbox{Partition-dependent} gate group modeling provides more accurate cost estimates than \mbox{hyperedge-based} approaches, which assign fixed costs regardless of partition.
Since gate group costs can vary significantly depending on which qubits are separated, our approach captures these savings and enables better partitioning decisions, particularly in circuits with diverse gate group constellations.

    \begin{table*}[ht]
    \centering
    \caption{Comparison of State-of-the-Art Algorithms vs. Proposed Work with and without Gate Groups (GG)}\label{tab:comparison}
    \vspace*{-2pt}
    \small
    \setlength{\tabcolsep}{2pt}
    \renewcommand{\arraystretch}{0.8}
    \begin{tabular}{l r r | r r | r r | r r | r r | r r}
        \toprule

        \multicolumn{3}{c|}{\textbf{Benchmark}} &
        \multicolumn{2}{c|}{\textbf{$\quad$MIP Gate~\parencite{gurobi}$\quad$}} &
        \multicolumn{2}{c|}{\textbf{CutQC Wire~\parencite{tang_cutqc_2021}}} &
        \multicolumn{2}{c|}{\textbf{$\qquad$Qiskit~\parencite{qiskit-addon-cutting}$\qquad$}} &
        \multicolumn{2}{c|}{\textbf{Prop. Work w/o GG}} &
        \multicolumn{2}{c}{\textbf{Prop. Work w/ GG}} \\[1pt]

        \hline
        
        Name & \thead[r]{Num.$\,$\\Qubits} & \thead{Num.\\2Q-Gates} &
        \thead{Runtime\\(ms)} & \thead{Sampling\\Overhead} &
        \thead{Runtime\\(ms)} & \thead{Sampling\\Overhead} &
        \thead{Runtime\\(ms)} & \thead{Sampling\\Overhead} &
        \thead{Runtime\\(ms)} & \thead{Sampling\\Overhead} &
        \thead{Runtime\\(ms)} & \thead{Sampling\\Overhead} \\[-2pt]

        \hline
        
        \rule{0pt}{9pt}QAOA & 8  & 16  & 1  & \sci{1.7}{3} & 1488  & \sci{6.5}{4} & 150  & \sci{1.7}{3}$\ $ & 123 & \sci{1.7}{3} & 260 & \sci{2.6}{2} \\
                            & 15 & 30  & 2  & \sci{1.7}{3} & 1581  & \sci{6.5}{4} & 1147 & \sci{1.7}{3}$\ $ & 200 & \sci{1.7}{3} & 303 & \sci{2.6}{2} \\
        QFT                 & 6  & 18  & 5  & \sci{2.5}{3} & 1584  & NA           & 149  & \sci{2.5}{3}$\ $ & 99 & \sci{2.5}{3} & 428 & \sci{4.2}{1} \\
                            & 12 & 72  & 14 & \sci{8.0}{5} & 49088 & NA           & 2365 & \sci{1.5}{7}$\ $ & 397 & \sci{8.0}{5} & \multicolumn{2}{c}{[Future Work]} \\
        Grover              & 4  & 38  & 1  & \sci{8.2}{6} & 1595  & NA           & 510  & \sci{1.7}{19} & 159 & \sci{8.2}{6} & 770 & \sci{5.1}{4}\\
        \multicolumn{2}{c}{Chain (\cref{fig:block-circuit})$\quad$10} & 36 & 3 & \sci{5.3}{5} & 1506 & \sci{1.6}{1} & 12 & \sci{1.6}{1}$\ $ & 235 & \sci{1.6}{1} & 257 & \sci{1.6}{1} \\
                                   & 20 & 73 & 4 & \sci{5.3}{5} & 1569 & \sci{1.6}{1} & 23  & \sci{1.6}{1}$\ $ & 429 & \sci{1.6}{1} & 561 & \sci{1.6}{1} \\
        \cref{fig:problem-example} & 6  & 28 & 2 & \sci{5.3}{5} & 1579 & NA           & 248 & \sci{2.1}{4}$\ $ & 153 & \sci{2.1}{4} & 212 & \sci{7.1}{3} \\[0.5pt]
        
        \hline
        
        \rule{0pt}{9pt}\textbf{Median} & & &
        \textbf{2.5} & $\bm{5.3\!\times\!10^{5}}$ &
        \textbf{1580} & $\bm{2.5\!\times\!10^{4}}$ &
        \textbf{199} & $\bm{2.1\!\times\!10^{3}}$ &
        \textbf{179.5} & $\bm{2.1\!\times\!10^{3}}$ &
        \textbf{303} & $\bm{2.6\!\times\!10^{2}}$ \\[-2pt]

        \bottomrule
    \end{tabular}
    \vspace*{-12pt}
\end{table*}

\section{Evaluation}\label{sec:evaluation}

The proposed method is the first to combine gate and wire cutting with joint cutting in a unified framework.
To demonstrate the effectiveness of the proposed method, we compared it with the state of the art in circuit cutting across various benchmarks.
The following sections describe the experimental setup and eventually present the obtained results.

\subsection{Experimental Setup and Metrics}

The proposed method has been implemented in Python 3.13 using NetworkX~\parencite{networkx}.
Evaluation employs the MQT Bench suite~\parencite{mqtbench} (QAOA instances for max-cut problems, QFT, and Grover circuits), plus synthetically generated benchmarks: chain-structured circuits (\cref{fig:block-circuit}) and circuits from the problem formulation (\cref{fig:problem-example}).
These circuits are restricted to single- or two-qubit gates and listed in~\cref{tab:comparison} together with their qubit number and two-qubit gate count. %
For each test scenario, we collect the runtime of the cutting method and the resulting sampling overhead.
Partitions are constrained to be at most \mbox{$3/4$} of the original qubit count, and we perform multiple runs (50 per instance) to achieve high-quality results.
Scalability is tested on an AMD Ryzen Threadripper PRO 5955WX (16 cores, 130GB RAM); tool comparisons use a MacBook Pro with Apple M5 (10 cores, 24GB RAM).

\subsection{Scalability}

\begin{figure}[ht]
    \vspace*{-4pt}
    \centering
    \begin{subfigure}[t]{0.585\linewidth}
        \centering
        \includegraphics[trim=7pt 8pt 8pt 8pt,clip,width=\linewidth]{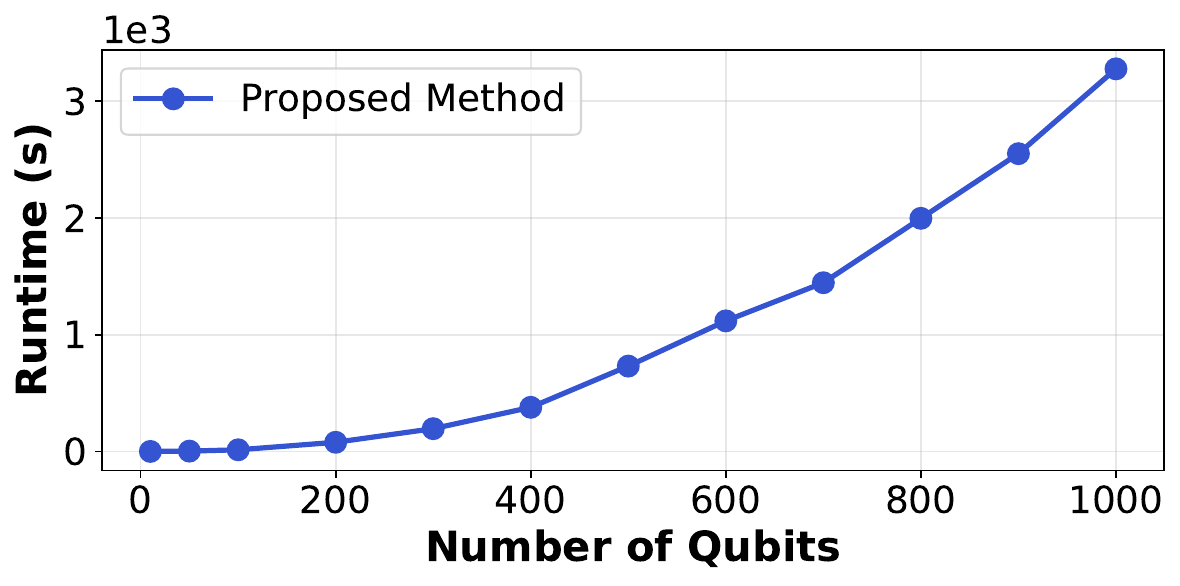}
        \caption{Scalability tests for random circuits.}\label{fig:scalability-tests}
    \end{subfigure}
    \begin{subfigure}[t]{0.4\linewidth}
        \centering
        \includegraphics[trim=2mm -4mm 0 0,clip,width=\linewidth]{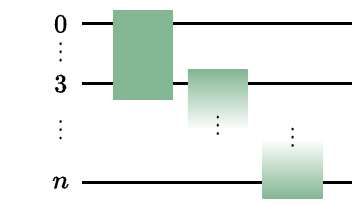}
        \caption{\emph{Chain Circuit} with $n$ qubits.}\label{fig:block-circuit}
    \end{subfigure}
    \caption{Scalability results and generated benchmark circuit.}\label{fig:evaluation}
    \vspace*{-8pt}
\end{figure}

To demonstrate scalability, we evaluated the proposed method on random circuits ranging from 10 to 1000 qubits with moderate gate density.
The method runs 50 times per instance, retaining the best solution.
The obtained runtimes, plotted in \cref{fig:scalability-tests}, scale moderately, making the proposed method suitable for assessing feasibility for circuit cutting across arbitrary problem instances.

\subsection{Comparison to Other Tools}

The proposed method was evaluated against three baselines: an MIP optimal gate-cutting approach using Gurobi~\parencite{gurobi}, CutQC~\parencite{tang_cutqc_2021} for optimal wire cutting, and Qiskit's circuit cutting implementation~\parencite{qiskit-addon-cutting} for both cut types.
The results are summarized in~\cref{tab:comparison}, with the middle columns showing the state-of-the-art baselines, and the rightmost showing the results of the proposed method with and without gate groups.

By combining gate and wire cuts, the approach minimizes or reduces overhead compared to either cut type alone.
This advantage is exemplified on the circuit in \cref{fig:problem-example}, where the method significantly reduces sampling overhead relative to the MIP baseline, underscoring the practical benefits of combined optimization.
Furthermore, while CutQC often fails to find valid solutions that satisfy the maximum qubit count, the proposed method consistently identifies viable cuts, demonstrating robustness across diverse circuit types.
Note that Qiskit performs k-way partitioning while the proposed method is restricted to bi-partitioning; appropriate qubit constraints ensure Qiskit returns only bi-partitions for fair comparison.
The proposed method occasionally incurs higher runtimes than Qiskit on some benchmarks.
Despite this trade-off, sampling overhead is comparable or superior across benchmarks, particularly for dense circuits such as QFT and Grover.

The last two columns in~\cref{tab:comparison} show the results when considering Gate Groups (GG).
The resulting runtime overhead is moderate and could be mitigated through caching strategies.
Automatic gate-group identification remains an open problem, with existing approaches limited to canonical forms or greedy heuristics~\parencite{andres-martinez_automated_2019,burt_generalised_2024}.
Hence, group identification currently is manual rather than automatic, limiting scalability for larger circuits.

Overall, the results obtained when considering gate groups demonstrate substantial performance improvements, especially on circuits with many parallel gates.
In particular, chain circuits achieve optimal sampling overhead that matches wire-cutting baselines, while dense circuits like QFT achieve an over $90\%$ reduction in overhead.
These results underscore the substantial potential of gate-group-enhanced circuit cutting and motivate future work on automation.

    \section{Conclusion}\label{sec:conclusions}

This work presents a unified, scalable approach for circuit cutting that simultaneously optimizes gate and wire cuts, while considering joint gate cutting\textemdash{}addressing a critical gap in existing methods, which often handle cutting strategies in isolation and do not exploit joint cuts. %
By leveraging graph partitioning heuristics in a two-stage strategy, the proposed approach never performs worse than gate cutting alone while achieving overhead reduction in many cases.

The method is robust, providing valid solutions across different circuit types, and scales to $1000$ qubits efficiently. %
Additionally, it provides diagnostic value: persistently high cut values reliably signal that a circuit is fundamentally unsuitable for cutting, enabling early go/no-go decisions in compilation pipelines.
Furthermore, gate group modeling for joint cutting significantly reduces overhead compared to prior approaches.
Together, these improvements make circuit cutting more practical for near-term quantum systems.

    \subsection*{Acknowledgments}\label{sec:ack}
    \footnotesize{\scalefont{1.15}
    We thank S. Volkamer for valuable feedback on the manuscript.
    The authors acknowledge funding from the European Research Council (ERC) under the European Union's Horizon 2020 research and innovation program under grant agreement No. 101001318, %
    as well as the Munich Quantum Valley (MQV), which is supported by the Bavarian state government with funds from the Hightech Agenda Bayern Plus. %
    Additionally, this work was supported by the BMFTR under grant number 13N17298 (SYNQ) %
    and by BMIMI, BMWET, the State of Upper Austria and the State of Tyrol within the COMET module Quantum Algorithm Engineering (FFG grant no. 923923) managed by the Austrian Research Promotion Agency FFG. %
    
    During the preparation of the code and manuscript, the authors used GitHub Copilot, powered by Claude's Haiku 4.5 to improve code, spelling, grammar, clarity, and readability.
    Afterward, the authors reviewed and edited the content as needed.
    The authors take full responsibility for the final content.
    }
    
    \renewcommand*{\bibfont}{\small} %
    \printbibliography%

\end{document}
\typeout{get arXiv to do 4 passes: Label(s) may have changed. Rerun}